# Tuning of quantum entanglement of a superconductor by Transition-metal and Rare-earth impurity effect and the role of potential scattering on quantum phase transition

N. Ebrahimian[1], M. Khosrojerdi[2], R. Afzali[2*]

[1]Department of Physics, Faculty of Basic Sciences, Shahed University, Tehran, 3319118651, Iran

[2]Department of Physics, K. N. Toosi University of Technology, Tehran, 15418, Iran

Emails: n.ebrahimian@shahed.ac.ir ,

 khosrojerdi@alumni.kntu.ac.ir,

*Corresponding author: afzali@kntu.ac.ir

## Abstract

By considering transition-metal (Shiba-Rusinov model) and rare-earth metal impurities (Abrikosov-Gor'kov theory) effect on a many-body system, i.e., a BCS s-wave superconductor, quantum bipartite entanglement of two electrons of the Cooper pairs in terms of the exchange interaction, $J$, the potential scattering, $V$ (contrary to expectations playing an important role), and the distance of two electron spins of the Cooper pair is calculated at zero temperature by using two-electron spin-space density matrix (Werner state). In transition-metal case, we found new quantum phase transitions (QPTs). The changes of $J$, which causes to have localized excited state, $V$ and the pair interaction (via energy gap) lead to the displacement of the QPTs (interactions act in the same direction, however sometimes the pair interaction causes the competition with other interactions), regardless of their effects on the value of concurrence. To have the turning point, which is a QPT point, by the reduction of $|J|$, the system doesn't need to have the large $V$. For non-magnetic and magnetic (rare-earth) impurity cases, the concurrence versus the distance and collision times is discussed for all finite and infinite Debye frequency. The quantum correlation, instability of the system and what's more important QPT can be tuned by impurity.

## Introduction

During the last years, it became evident that the quantum entanglement (QE) is one of the most important resource in quantum information (QI) and quantum computation, especially, the use of the extraction of the QE in a many-body system. The extensive research efforts on QE were revealed both its qualitative and quantitative aspects of QI[1-6]. The method of the measuring of the entanglement is a broad field of research in its own. The consideration of the QI and QE in the many-body systems causes to reveal of the new properties[7]. It should be noted that in the many-body systems, entanglement is much more complex than other systems, and also, multipartite entanglement influences in the ground state at zero temperature. The generation and manipulation of the bipartite and multipartite entangled states through many-body Hamiltonians, such as the quantum computer's Hamiltonian were investigated[7-12]. Other investigations on many body systems such as the entanglement of electron spins are as follows. The entanglement of two electron spins of a noninteracting electron gas based on the Green's function approach was discussed[13-14]; also, multipartite entanglement in a non-interacting Fermi gas was studied[15]. Furthermore, few researches have been done on superconductors using space-spin density matrix approach. One of them is about bipartite entanglement of two electron spins forming Cooper pairs in a BCS s-wave superconductor[8]; another one is about bipartite and tripartite entanglement and quantum correlation of s-wave and d-wave bulk and nano grain superconductors, which was given by some of our authors[9,10].

It was found in numerous works that the entanglement, referring to quantum correlations between subsystems can be a good indicator of QPTs, which can be found in the many body systems[4,16]. A lot of work has revealed that the bipartite or the pairwise entanglement itself or its derivatives display local extremes close to quantum critical points (QCPs). It should be noted that generally the investigation of the entanglement itself is not necessary and sufficient condition for the occurrence of QPT; necessarily one-to-one correspondence between QPT and the appearance of the critical point on concurrence does not exist, unless under some conditions, which was usually used in some spin model such as Isingmodel[16-19]. However, one-to-one correspondence between QPT (with accompanied to discontinuities of the ground state) and the behavior of the matrix elements of density matrix always exist. Nevertheless, in our present work there is one-to-one correspondence between QPT and entanglement.

Here, we bring about some theories related to the impurity effect on superconductors. Over the past decades, researchers and scientists have paid to the investigation of the impurities effect on the properties of the superconductors[20-22]. Shiba, Rusinov, Abrikosov, and Gor'kov have provided theories about the Green's functions and some physical properties of the superconductors in the presence of two kinds of the impurities[23-27]. Shiba and Rusinov independently, have given the theory about the presence of the low concentration of the uncorrelated transition-metal impurities into a superconductor. In Shiba-Rusinov (SR) model, the scattering is calculated exactly for a single impurity problem by treating the impurity spin classically and it is shown that there exists a localized excited state in the energy gap[25-27]. The use of the SR model on the isotropic s-wave superconductors doped with the magnetic atoms is identified that the information on potential scattering and exchange interaction cannot be separately obtained; also, the energy of bound states is a function of them[28]. Also, the effect of the impurities on the anisotropic superconductors, Josephson current and several other properties of the superconducting alloy was studied by SR approach[29-33]. In the several valuable theoretical and experimental articles by using the lattice model, Bogoliubov-de-Gennes Hamiltonian, the Zeeman field and Majorana fermion approach, have been addressed to the investigation of the QPT in the superconductors in the presence of the magnetic impurity[34-40]. The robustness to the non-magnetic impurities of the isotropic s-wave superconductors based on the symmetry of the system approach was studied by Anderson[23]. The role of the rare-earth metal impurities case in the conventional superconductors was studied by Abrikosov and Gor'kov[24]. The Markowitz and Kadanoff anisotropic superconductivity is weakly suppressed by the paramagnetic impurity scattering effect[41].

The purpose of this paper is to investigate of QPTs (based on our knowledge, these are new QPTs) and the bipartite entanglement properties of an BCS s-wave superconductor, in the presence of the non-magnetic impurity and magnetic impurities (transition-metal accompanied with SR model and the rare-earth metal accompanied with AG theory) at zero temperature by using the space-spin density matrix, up to the first-order approximation. Meanwhile, the role of the potential scattering, which is exactly the same important role that than exchange interaction, on the competition among interactions is investigated. It merit mentioning that the increase of the absolute value of the exchange interaction in the occurrence of QPTs always causes the increase of the potential scattering and the effects of $|J|$ and $V$ in the concurrence are always coupled. The values of the all available parameters such as the coefficients of the interactions strongly depend on the assumptions, which are tuned by the numerical calculations. The assumptions are first, the smallness of the perturbed (in the presence of the impurity) Green's functions in comparison with the unperturbed Green's functions ( in all distance of the electrons of the Cooper pair ) and second, the smallness of the appeared parts due to the impurities in the renormalized order parameter (or energy gap), and the single-particle energy and frequency. First, we address the investigation of the role of the exchange interaction, potential scattering, and normalized transition-metal impurity concentration in the SR model by considering the entanglement approach. Furthermore, the appearance, the determination and the displacement of QPTs based on the values of the parameters especially potential scattering regardless of the influence of the parameters on the value of concurrence, are investigated. We show that potential scattering plays an important role in the SR model contrary to expectations. Second, We use the results of the AG theory for the rare-earth metal (except cerium) and non-magnetic impurities in the superconductors up to the first order approximation. One of the most important assumptions in the AG theory is that the exchange interaction between a conduction electron and a magnetic impurity spin is weak and the lowest-order Born approximation is used to treat the scattering. For both rare-earth metal and non-magnetic impurity, we pay attention to investigate both infinite and finite Debye frequencies; it should be mentioned that previously, Green's functions were considered with infinite Debye frequency[21]. Furthermore, the comparison among the finite Debye frequency, the order parameter (energy gap) and the frequency gives different conditions on their appropriate values for the numerical analysis of concurrence.

## Results and discussion

### A. The transition-metal impurity case

First, we consider the effect of the transition-metal impurity, which is described by the SR model, on the s-wave superconductor, by taking care of the critical concentration impurity and considering the potential scattering that has an important role to quantum correlation such as QE. It merits mentioning that unexpectedly, not only the potential scattering as well as the exchange interaction plays the important role in concurrence, but also, the existence of the QPTs depends on the presence of the potential scattering. Furthermore, of course, if exchange interaction is nonzero, then, the potential scattering becomes important term; otherwise, the potential scattering acts like non-magnetic impurity in the AG theory. The concurrence $(c \equiv c\,(p,j,v,k_F r)\,)$ in the presence of the impurity is calculated as a function of the external and internal parameters, i.e., $v$, which is related to the potential scattering, $V$, via $V \equiv v/\pi N(0)$, $j$, which is defined by the exchange interaction via $J \equiv 2j/S\pi N(0)$, $p$, which is defined as the normalized impurity concentration, and $k_F r$. Meanwhile, $N(0)$, $S$, $r$ and $k_F$ are the density of single-particle states at the Fermi surface, the spin of the impurities, the distance of two electrons of the Cooper pairs and Fermi wave number, respectively. We study of the effect of the variations of these parameters on concurrence and QPTs for the transition-metal impurity case (throughout the paper, we use $\hbar = 1$ ).

Concurrence as a measure of bipartite entanglement in the presence of the transition-metal impurities versus $v$ at different fixed values of $p$, $k_F r$, and $j$ are depicted in Fig.1. We would like to consider Green's functions up to the first-order approximation to calculate concurrence, so that, the ratio of the perturbed Green's functions to the primary Green's functions must be always controlled to have a small value. Therefore, by decreasing the value of $j$, the appropriate range of the values of $v$ becomes more limited to satisfy our approximation, so that for any step of our calculation, the situation must be checked. This issue can be usually achieved by considering the higher values of $|j|$ and the lower value of $p$. Also, it should be mentioned that different fixed values of $k_F r$ will not affect on the restrictions over the values of $v$ to meet our approximation. The existence and appearance of the critical points (such as the turning point and the local maximum point) depend on the values of $j$. Of course, to appear the turning point and the local maximum point, we have chosen some special values of $j$. Also, it should be mentioned that the only effect of different fixed values of $p$ is the change of the value of concurrence but the overview of the concurrence does not change. As it can be seen from Fig.1, the concurrence increases with increasing the normalized impurity concentration at the fixed value of $v$. In the insertion of Fig.1, the concurrence versus $v$ in the presence of transition-metal impurities for different fixed values of $k_F r$ at a fixed $p$ and a fixed $j$. As expected, by increasing the relative distance between two electrons of a Cooper pair of a conventional superconductor (or an s-wave superconductor), the concurrence reduces. The behavior of the curves of the insertion is similar to each other and all critical points occur at the same value of $v$. The curves in Fig.1 change from concave down to concave up that this occurrence produces a turning point. For each $v$ that located in the almost interval (0.70,0.95) with $|j| = 0.8$, curves are completely ascending.

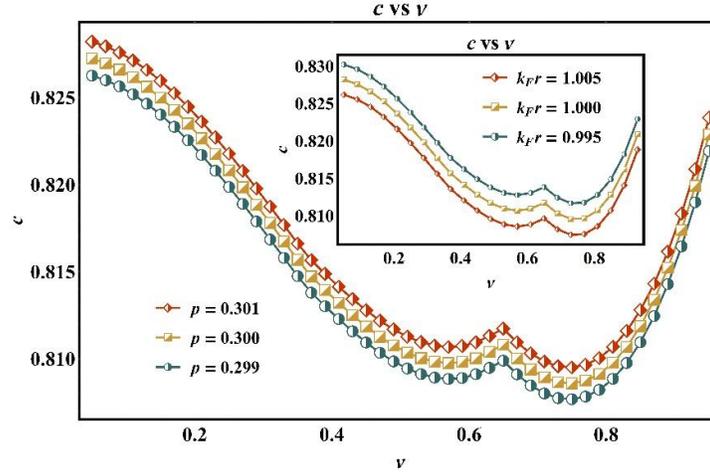

**Figure 1.** For transition-metal case; concurrence versus $v$ for various values of the $p$ at $j = -0.8$ and $k_F r = 1$. Inset: concurrence versus $v$ for various values of the $k_F r$ at $j = -0.8$ and $p = 0.3$.

Concurrence versus $v$ at different fixed values of $j$, at a fixed value of the relative distance between two electrons of a Cooper pair of an s-wave superconductor and a fixed value of the normalized impurity concentration, is depicted in Fig.2. The curves existing in Fig.2 are intentionally brought, because of showing how to disappear one of the QCP (local maximum point) with reducing the absolute value of the exchange interaction; by using derivatives of concurrence, we find that the annihilation of the critical point occurs at about $j \approx -0.60$ with given typical energy gap and Debye frequency. It is worth mentioning that, by increasing the value of $p$, the appropriate range of the value of $v$ becomes more restricted to satisfy our approximation (the values of $p$ and $k_F r$ were typically chosen by 0.2 and 1, respectively). From Fig.2, by changing the value of $j$, the concurrence and bipartite entanglement is changed; the QCPs have been formed when the values of $j$ has been located almost between -0.60 and -0.99. It is seem that by reducing the value of $|j|$, these points are displaced into the left-hand side of the curves. For each $j$ located in the almost interval$(-0.05, -0.60)$, such as the curve identified with $j \approx -0.60$ in Fig.2, the curve does not show any critical point. For $j = -0.68$, in the interval$v \approx (0.05, 0.4)$, the variation of concurrence wasn't noticeable, but after passing the local maximum (minimum) point, concurrence decreases (increases) up to $v \approx 0.6$ ($v \approx 0.95$). By increasing the absolute value of $j$, both the turning point and the local maximum point move to the higher values of $v$ (move to the right-hand side on the $v$ axis). The turning point (the local maximum point)is disappeared at$j < |-0.64|$ ($j < |-0.60|$). Of course, it should be mentioned that this disappearance is influenced by the approximation used in all calculations and the range used for the potential scattering. In addition, it can be seen that by increasing the absolute value of $j$, the concurrence decreases.

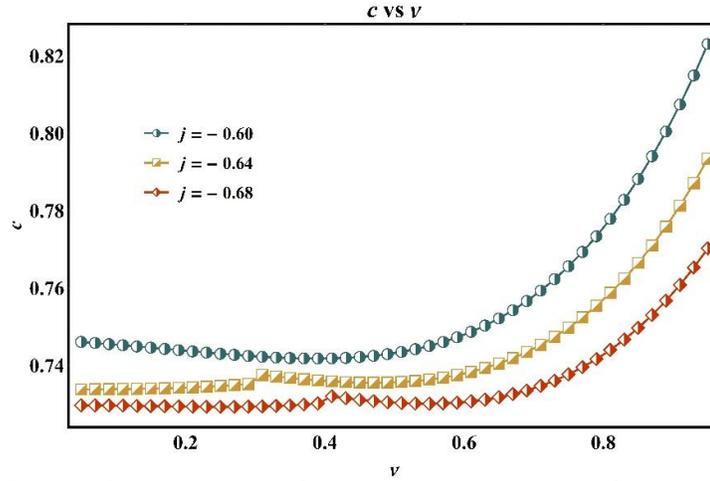

**Figure 2.** For transition-metal case; concurrence versus $v$ for various values of the $j$ at $p = 0.2$ and $k_F r = 1$.

It should be noted that in the presence of the magnetic impurity, the energy gap is reduced[25]; this reduction depends on the normalized impurity concentration, the exchange interaction, the potential scattering and density of states. Our calculation shows that the change of the energy gap due to the magnetic impurity is very small, so that we can neglect the effect of this reduction on the energy gap and thereby, the maximum error occurred in the calculation is about few percent. Therefore, we have used a typical energy gap such as that than given in Ref. [8]. However, for s-wave superconductors with different energy gap (and also Debye frequency), we can show that how the change of the energy gap can affect not only the value of concurrence but also the occurrence of the critical point at different value of the exchange and potential scattering. Concurrence versus $v$at different fixed values of the $\Delta$ is depicted in Fig.3. At different values of $\Delta$ or pairing interaction, the turning point and the local maximum point can be displaced and these points can be disappeared up to the first-order approximation; for example, by considering the selected values for $\Delta$ in Fig.3, the lowest curve doesn't show any the maximum local point. By increasing the value of $\Delta$, the turning point (the local maximum point) move toward the lower (higher) value of the $v$. As well, as it can be seen from Fig.3, the concurrence increases by increasing the value of $\Delta$.

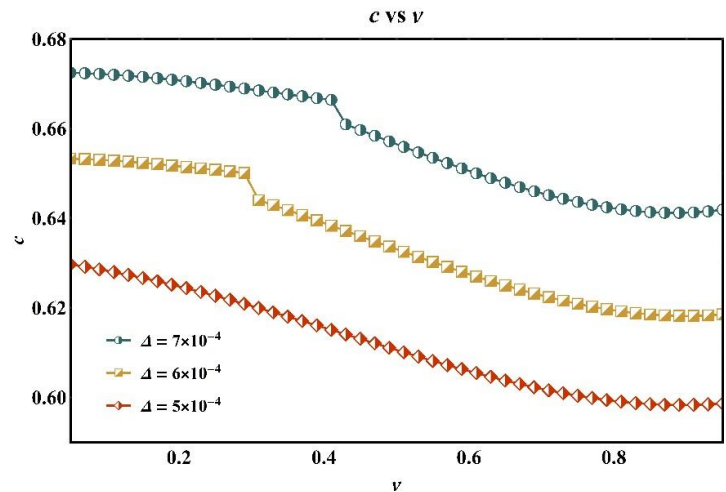

**Figure 3.** For transition metal case; concurrence versus $v$ for various values of the $\Delta$ at $j = -0.8$, $p = 0.2$ and $k_F r = 1$.

The study of the partial derivative of concurrence versus $v$, by considering the typical case of $j = -0.8$, $k_F r = 1$, and $p = 0.3$, gives us valuable information about the critical points such as turning point and local maximum point. In Fig.4(a), for each $v$ that located in the almost interval $(0.1, 0.5)$, at $v \approx 0.29$, the first partial derivative of $c$ versus $v$ has a minimum, the value of $\partial_v^2 c$ is equal to zero and the value of the third partial derivative wasn't zero. Concurrence is continuous at a point $v \approx 0.29$ and also at this point, which is the turning point, the curve is changed from concave down to concave up. In addition, from Fig.4 (b), for each $v$ that located in the almost interval $(0.635, 0.665)$, at $v \approx 0.650$, $\partial_v c$ is equal zero and the second partial derivative has a minimum, and it can be seen that this point, which is the local maximum point, behaves like as a critical point. we address the investigation of the curves of $c$ vs $v$ in three regimes, at a fixed exchange interaction, from lower to the higher value of the potential scattering, the first regime appears up to receiving the turning point (first the concurrence reduces with the sharp slope, then the slope of the curves becomes reverse), the second regime is started from this point and is contained the local maximum point, finally, the third regime is included the curves that are quite ascending with respect to potential scattering. Analytical and numerical studies on the concurrence (also via energy of the system) and its first and second derivatives sign to have QPTs. By changing the values of $j$ and $\Delta$, these critical points (turning point and local maximum point) occur at different $v$; also, the normalized impurity concentration and the relative distance between two electrons of a Cooper pair of a conventional superconductor do not change the value of $v$ of QPT points, which will be discussed in the following.

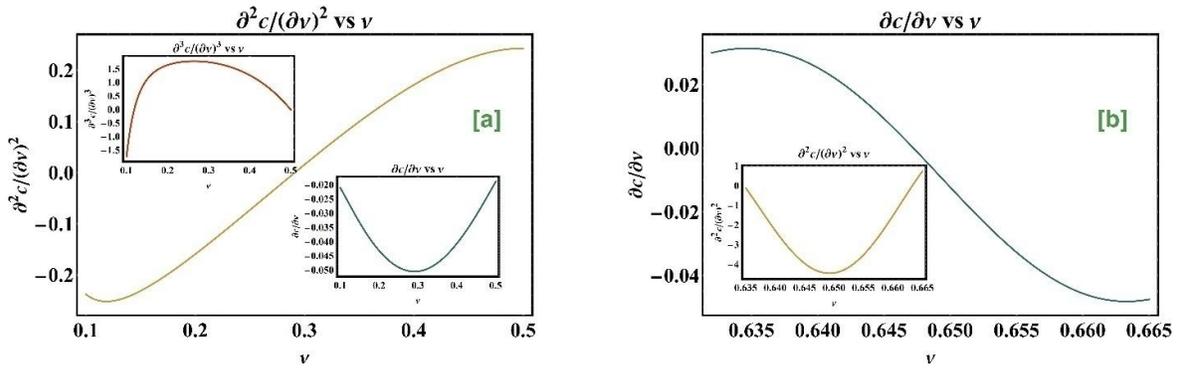

**Figure 4.** partial derivatives of $c$ vs $v$ at $j = -0.8$, $k_F r = 1$ $and$ $p = 0.3$ **(a):** for each $v$ that located in the interval $(0.1, 0.5)$ **(b):** for each $v$ that located in the interval $(0.635, 0.665)$.

Now, we proceed to interpret the change of the bipartite QE of two electron spins forming Cooper pairs. It should be mentioned that the concurrence (also via energy of system) shows the existence of QPTs; there are the necessary and sufficient conditions for QCPs existing in Figs. 1 and 2 to become QPT points; we exactly investigate the behavior of these points also by the change of energy of system, and its first and second derivatives. There is one-to-one correspondence between QPT and entanglement. The turning point and the local maximum point produced due to the potential scattering are new QPT points. It worth mentioning that when the cooper pair exists and bipartite entanglement is nonzero, the relative distance between the electrons of the Cooper pair is not influenced to displace the characteristics of the QPT points such as $j$ and $v$. Previously, at the point $j = 0$, where the system goes from classical spins model to quantum spin model, the system has been shown a QPT; also, the point $j = -1$, where the bound state is induced, is a QPT point[25-27].However, our investigation about QPTs via QE is related to the value of $j$ located between 0 and 1with taking into account the potential scattering (the increase of the $v$, corresponding to the increase of the absolute value of $j$) up to the first-order approximation. It is worth mentioning that we know exactly what happens in the system and how we can see the interaction influences on the electron spins' correlation of the Cooper pair. In the other words, it is very interesting that to follow the effect of the all real interaction (and not the manipulating the virtual interaction in the virtual system) in the real interesting many-body system, i.e., the interacting Fermi system, so-called the superconductor. At a fixed potential scattering, the concurrence in terms of the potential scattering, which was usually not considered in the investigation of the physical quantities in the SR model, shows that by the reduction of the absolute value of the exchange interaction (the effect of the localized excited state increases by getting stronger the coupling the impurity with conduction electrons before the induced impurity band is formed) or the increase of the interaction between Cooper pair via the energy gap, the value of concurrence is increased. Hence, the competition of these interactions occurs in the concurrence. It is worth mentioning that at the turning point, regardless of the effect of the interactions on the competition of the value of concurrence, there is the competition between of the intensities of the exchange and pair interaction to displace the turning point to lower or higher value of the potential scattering; the enhancement of the absolute value of the exchange (pair) interaction tends to displace the point to the higher (lower) value. In other words, in order to have the turning point, by the increase of the pair interaction (or the reduction of the absolute value of the exchange interaction), the system does not need to have a large potential scattering. Another important note is that the increase of the absolute value of the exchange and pair interaction causes the local maximum point is occurred at the higher values of the potential scattering. It should be noted that the effect of these interactions on the value of concurrence at the local maximum point is reversed; i.e., the increase of the absolute value of the exchange (pair) interaction causes the reduction (increase) of the concurrence at the local maximum point. It is worth mentioning that the increase of $|J|$ in the occurrence of QPTs always causes the increase of $V$ and the effects of $|J|$ and $V$ in the concurrence are always coupled.

Concurrence versus $j$ at different fixed values of $v$ and at the fixed values of $p$ and

$k_F r$, is depicted in Fig.5. The behavior of the curves are similar, so that by increasing the absolute value of $j$, the local maximum appears on the concurrence. To investigate of the curves of concurrence versus $j$, it is better that two regimes on the $j$ axis are considered. For each $j$ located in the almost interval$(-0.30, -0.60)$, by increasing the value of $v$, the concurrence increases, whereas, for each $j$ located in the almost interval$(-0.60, -0.99)$, the situation is reverse. It is worth mentioning that, after lengthy but straightforward calculations with different fixed values of $p$ and $k_F r$, it is observed that an overview of the curves doesn't change.

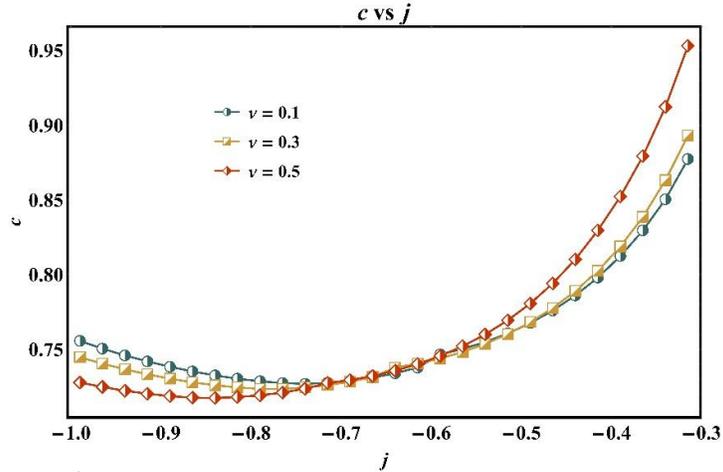

**Figure 5.** for transition-metal case; concurrence versus $j$ for various values of the $v$ at $p = 0.2$ and $k_F r = 1$.

By considering the fixed value of $v$ (= 0.1), the relative distance between two electrons of a Cooper pair of the s-wave superconductor, $r$ (= $1/k_F$), and a fixed normalized impurity concentration, $p$ (= 0.2), the first and second partial derivatives of the concurrence versus $j$ are depicted in Fig.6. It can be seen that at $j \approx -0.94$, $\partial_j c$ has a local minimum and the second partial derivative is equal zero.

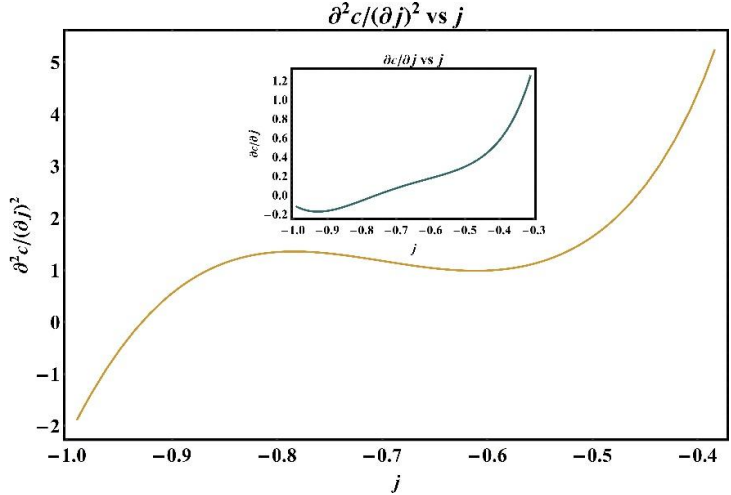

**Figure 6.** partial derivatives of $c\ vs\ j$ at $v = 0.1, k_F r = 1\ and\ p = 0.2$.

As it can be seen from Fig.5 and 6, there is a turning point in the curve of $c$ versus $j$. More importantly, at different potential scattering, the comparison of curves indicates to have different behavior on concurrence.

Concurrence versus the relative distance between two electrons of a Cooper pair at different fixed values of $p, v$, and $j$ is depicted in Fig.7. As it can be seen that from the overview of the figures, by increasing the value of $k_F r$, the concurrence decreases; what's more important, at a fixed the value of $k_F r$, the value of concurrence in the presence of the transition-metal impurity is greater than that of the non-impurity case, at the fixed other parameters. From Fig.7(a), it can be seen that the curves for different values of $v$, almost overlapped. From Fig.7(b), it can be seen that at a fixed value of $k_F r$, by increasing the value of $p$, the concurrence is increased. At the higher value of $p$, the zero value of concurrence occurs at larger value of $k_F r$. Finally, from Fig.7(c), it can be seen that by increasing the absolute value of $j$, at a fixed $k_F r$, concurrence decreases. By considering the high absolute value of $j$, the concurrence almost doesn't change. Also, up to the first-order approximation, by changing the $k_F r$, at the fixed other parameters, the value of concurrence changes, but it is not affected on the QPT points identified by $(j, v)$.

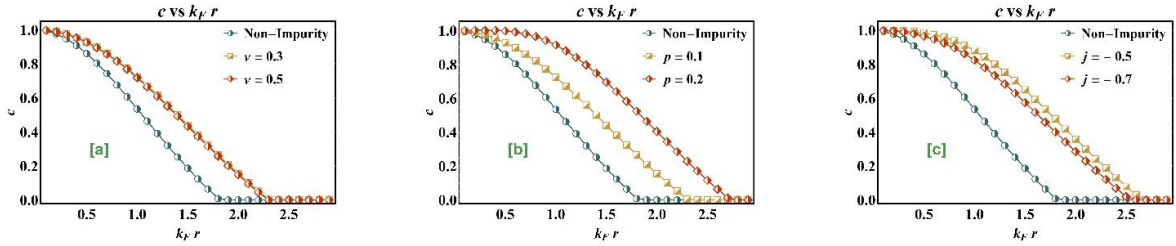

**Figure 7.(a):** for transition-metal case; concurrence versus $k_F r$ for various values of $v$ at $j = -0.8$ and $p = 0.2$. **(b):** concurrence versus $k_F r$ for various values of $p$ at $j = -0.3$ and $v = 0.3$. **(c):** concurrence versus $k_F r$ for various values of $j$ at $p = 0.3$ and $v = 0.3$.

### B. Non-magnetic impurity case

Detailed and comprehensive investigation over the influence of the non-magnetic impurities on concurrence is done by using AG theory after straightforward lengthy calculations, in which Green's functions are calculated by considering the different situations of the Debye frequency. It should be mentioned that the value of the Debye frequency is finite for real materials unless the Debye frequency is very high or is supposed to have infinite value for simplicity purposes[21]. Usually, Green's functions in the presence of the impurity were considered at the infinite value of the Debye frequency, whereas, we consider both infinite and finite cases. When we use the Debye frequency with infinite (finite) value, Green's functions are so-called as $G_{NM_1}(G_{NM_2}, G_{NM_3}$ and $G_{NM_4})$. It can be seen that from the overall view of Fig.8 (a), at a fixed $\tau$ ($= 10$), by increasing the value of $k_F r$, concurrence decreases. For the purpose of comparison, we bring concurrence without any impurity[8,9]. The curves of the concurrence in the presence of the non-magnetic impurity with finite $\omega_D$ case and non-impurity case are overlapped, however, at a fixed value of $k_F r$, for the infinite Debye frequency case, concurrence is less than other cases. In Fig.8 (b), we focus on the investigation of the concurrence versus inverse collision time at a fixed value of $k_F r$ ($= 1 \times 10^{-3}$). By increasing the value of the inverse collision time, for the infinite (finite) Debye frequency case, the concurrence decreases linearly (doesn't change significantly). For a fixed $1/\tau$, for the infinite Debye frequency case, the concurrence is smaller than that of the finite Debye frequency case.

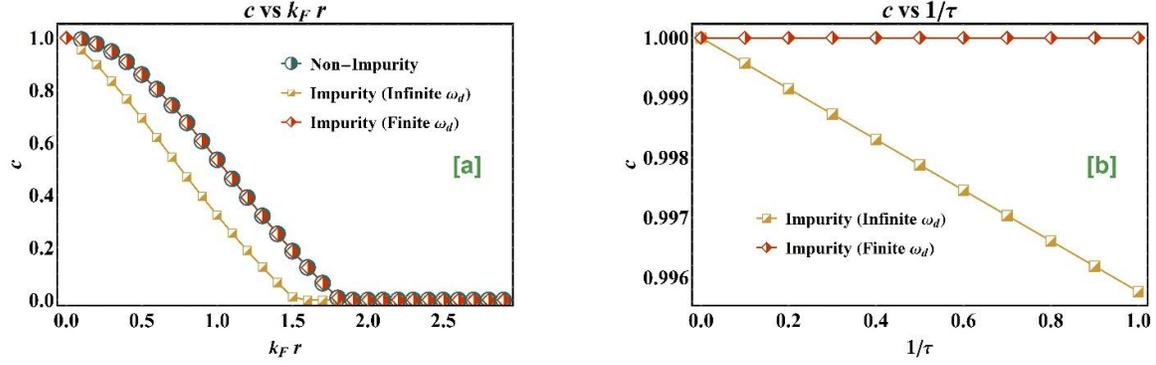

**Figure 8.(a):** for Rare-earth metal case; concurrence versus $k_F r$ for infinite and finite Debye frequency in the presence of the impurity and non-impurity cases. **(b):** concurrence versus inverse collision time for infinite and finite Debye frequency in the presence of the impurity and non-impurity cases.

### C. The Rare-earth metal impurity case

The investigation of the concurrence in the presence of the Rare-earth impurity versus $k_F r$, $\tau_1$, and $\tau_2$ for both the infinite and finite Debye frequency cases is done. The appropriate range of the $\tau_1$ and $\tau_2$ are tuned by considering the kinds of the Debye frequency up to the first-order approximation. After long, comprehensive and very massive calculations, we result that the value of the $\tau_1$ and $\tau_2$ must be very long (the order of both the $\tau_1$ and $\tau_2$ for the infinite (finite) Debye frequency is about $10^{25} s$ ($10^{39} s$) without physical meaning) to satisfy our approximation.

## Methods

The entanglement gives rise from two ways; one of them is particle statistics, so the noninteracting system shows entanglement, another way, which leads to an increase or decrease of the bipartite or tripartite or even multipartite entanglement, is due to external or internal interaction, which is found in the few- or many- body systems[9]. For the calculation of the reduced space-spin density matrix and required Green's functions of our system, which is an interacting Fermi gas, we have considered the trick that causes the procedure calculating of the concurrence in the noninteracting systems can be used for the interacting systems. Previously, for s-wave and d-wave superconductors as the interacting systems, the trick was used and the only change was to apply Green's functions for superconductors instead of noninteracting Fermi gases[8-9,13]. The reason of the extraction and the use of the formulation of the interacting system based on the noninteracting system can be explained from other viewpoints as follows; first, a superconducting ground state can be writing as the product of the ground state of a noninteracting Fermi system to an extended Jastrow function or Feenberg factor[9,42-43], which is related to the interactions and caused to modify bipartite entanglement in the noninteracting part of the state. Second, by Bogoliubov transformation, Hamiltonian of the system can be diagonal as a noninteracting system but with different energy[9]. Of course, the existence of impurity on superconductors will produce additional interaction, which can be considered as a small perturbation on superconductors. Finally, Green's functions of the system can be written similarly to that those of the noninteracting system, like as the ground state of the interacting system under investigation, i.e., the superconductor in the presence of the impurity, which is written via that those of the noninteracting system. Thereby, the formulation of the concurrence in terms of the noninteracting Green's functions can be survived. It should be mentioned that at least the new Green's functions (in the presence of the impurity) has the same form as the absence of the impurity, but only by transforming such as $\omega$ to $\widetilde{\omega}$. Another important point is about the structure of the BCS theory, in which mean-field approximation is used. We know that this approximation can change the entanglement so that the results of bipartite entanglement maybe incorrect unless fluctuation related to all interactions can be considered to be small (in all paper, smallness of the fluctuation is considered). By these assumptions and explanation, first, we bring the relation between reduced space-spin density matrix (and its elements are written in the spin coordinate and each of the elements depends on the space coordinate) and Green's functions (fermionic case) of a superconductor. We ignore anomalous Green's function of the system, because unperturbed anomalous Green's function in comparison to Green's function can be neglected and according to that, the small modification due to the impurity on anomalous Green's function becomes so small. Then, we have

$$\rho^{(2)}(\vec{x}_1, \vec{x}_2, \vec{x}_1', \vec{x}_2') = (-1/2)(G(\vec{x}_1 t_1, \vec{x}_1' t_1^+)G(\vec{x}_2 t_2, \vec{x}_2' t_2^+) - G(\vec{x}_1 t_1, \vec{x}_2' t_2^+)G(\vec{x}_2 t_2, \vec{x}_1' t_1^+)) \quad (1)$$

By using normalized reduced space-spin density matrix, which is again the Werner state as a superconductor in the absence of the impurity, and considering the case $\vec{x}_1 = \vec{x}_1'$, $\vec{x}_2 = \vec{x}_2'$ and $\vec{r} = \vec{x}_1 - \vec{x}_2$, then we have concurrence, as measure of quantum bipartite entanglement[3,14], $c = max\,[0, (3(G(r)G(-r)/(2 - G(r)G(-r))) - 1)/2]$. In $G$, the dependence of angles was done and concurrence can be linearized with respect to the all appropriate variables up to first-order approximation. Now, for this purpose, we proceed to bring the calculated Green's functions.

**Transition-metal case.** We considered the transition-metal impurity as a perturbation in Green's functions, thereby, in the reduced density matrix and concurrence of the system. The ratio of the perturbed Green's functions to the primary Green's functions is tuned at each step of the calculation up to the first-order approximation, so that its value becomes small. This approach limits the range of the parameters (like the exchange interaction, the potential scattering, and the renormalized impurity concentration) existing in the concurrence, which has been considered in the numerical calculation by choosing the value of each parameter and then fixing other parameters to meet the above condition. We start to write the exact Green's functions[25-28], with the following notation

$$\bar{G}_{\vec{k}} = \begin{pmatrix} a_1 + a_2 & 0 & 0 & -a_3 \\ 0 & a_1 + a_2 & a_3 & 0 \\ 0 & a_3 & a_1 - a_2 & 0 \\ -a_3 & 0 & 0 & a_1 - a_2 \end{pmatrix} \quad (2)$$

where $a_{n(=1,2,3)} = \tilde{b}_n / (\tilde{\omega}^2 - \tilde{\Delta}^2 - \tilde{\epsilon}_{\vec{k}}^2)$, $\tilde{b}_1 = \tilde{\omega}$, $\tilde{b}_2 = \tilde{\epsilon}_{\vec{k}}$ and $\tilde{b}_3 = \tilde{\Delta}$. Also, $\tilde{\omega}$, $\tilde{\epsilon}_{\vec{k}}$, and $\tilde{\Delta}$ are the renormalized frequency, renormalized kinetic energy with respect to chemical potential, and renormalized order parameter, respectively. It should be mentioned that $\tilde{b}_n = b_n + \Gamma_n \times \zeta_n(j, v, U(\tau_s, \omega/\Delta, j, v))$ where $\tau_s$ is collisiontime and $\Gamma_n = p\Delta t_n(j,v)/4j^2$. By considering that the function $t_n(j,v)$ is always less than one and $0 < p < 1$, then, $\Gamma_n$ can be chosen to be less than one, so in $\tilde{b}_n$, the additional term to $b_n$, which is due to the existence of the impurity, can be considered to be small and therefore, we would like to do the calculation up to first-order approximation. However, in the numerical calculation, we aware to check all quantities for satisfying our approximation, for example a part range of $p$ is acceptable. It should be noted that $\zeta_n$ can be approximated, when $-1 < j < 0, 0 < v < 1$, $\tau_s \ll 1$ and $U(\tau_s, \omega/\Delta, j, v) \approx \omega/\Delta < 1$. The perturbed Green's functions as follows:

$$\bar{G}^{(1)}\left(\epsilon_{\vec{k}}, \omega\right) = C\left(\epsilon_{\vec{k}}, \omega\right) \left(\Gamma_1 \omega \left(\left(\omega + \epsilon_{\vec{k}}\right)^2 + \Delta^2\right) - 2\Gamma_2 \Delta^2 \left(\omega + \epsilon_{\vec{k}}\right) - \Gamma_3 (\Delta^2 - \omega^2)^{1/2} \left(\left(\omega + \epsilon_{\vec{k}}\right)^2 - \Delta^2\right)\right) \quad (3)$$

where

$$C\left(\epsilon_{\vec{k}}, \omega\right) = \frac{(\Delta^2 - \omega^2)^{1/2}}{(\omega^2 - \Delta^2 - \epsilon_{\vec{k}}^2)^2 (\omega^2 - \Delta^2 \epsilon_0^2)} \quad (4)$$

Then, for $G(r)$, after integration on angles, we have

$$G(r) = \frac{(2\pi)^{5/2} N(0)}{r\, p_F} \int_{-\omega_D}^{\omega_D} d\epsilon \int_{-\Delta}^{\Delta} d\omega\, \tilde{G}\left(\epsilon_{\vec{k}}, \omega\right) \sin\left(p_F r + \epsilon_{\vec{k}} r/v_F\right) \tag{5}$$

According to the numerical calculation, the effect of the variation of the Debye frequency on the results is insignificantly, so, we choose $0.1 eV$ as the typical value of the Debye frequency. The value of order parameter or energy gap is measured with respect to the Fermi energy (we choose $\epsilon_F = 1\ eV$) and as said before, the energy gap is considered to be independent of the variation of the impurity concentration and we choose $0.001 eV$ as the typical value of the order parameter (Of course, we have investigated the different values of the order parameter).

**Rare-earth metal impurity and nonmagnetic impurity cases.** We investigate both cases based on AG theory. The Hamiltonian of the conventional superconductor in *BCS* theory in the presence of the impurity atoms is given by $H = H_0 + H_{int}$ where $H_{int} = \sum_a \int U(\vec{r} - \vec{r}_a) \psi_\alpha^\dagger(\vec{r}) \psi_\alpha(\vec{r}) d^3\vec{r}$. It is supposed that there are no correlations between different impurities. Here, we use the Green's functions $G(p, \omega) = \left(\omega \eta_\omega + \tilde{\epsilon}_{\vec{k}}\right)/\left(\omega^2 \eta_\omega^2 - \tilde{\epsilon}_{\vec{k}}^2 - |\Delta \eta_\omega|^2\right)$. Investigation of the Debye frequency in finite and infinite ranges is interested for us. The $\eta_\omega$ in two general separate conditions, i.e., $\omega^2 > \Delta^2$ and $\omega^2 < \Delta^2$ is given as follows:

$$\eta_\omega = 1 + \frac{1}{\pi\tau} \frac{ArcTan\left[\frac{\omega_D}{\sqrt{\Delta^2 - \omega^2}}\right]}{\sqrt{\Delta^2 - \omega^2}}, \qquad \Delta^2 > \omega^2 \tag{6}$$

$$\eta_\omega = 1 - \frac{1}{\pi\tau} \frac{ArcTanh\left[\frac{\omega_D}{\sqrt{\omega^2 - \Delta^2}}\right]}{\sqrt{\omega^2 - \Delta^2}}, \qquad \omega^2 > \Delta^2 \tag{7}$$

Infinite Debye frequency case:

At zero temperature, up to the first-order approximation, the Green's function in terms of the non-impurity Green's function is obtained by

$$\lim_{t \to 0^-} G_{NM_1}^{(1)}(\vec{r}, t) = \left(\lim_{t \to 0^-} \int G_0(r, \omega) e^{i\omega t}\, d\omega\right) e^{\frac{-r}{2\tau v_F}} \equiv (G_0(\vec{r})) e^{\frac{-r}{2\tau v_F}} \tag{8}$$

where $G_0(\vec{r})$ was given in Refs. [8-9]. Also, by considering the rare-earth metal impurity, up to the first-order approximation, the Green's functions are obtained as follows ($a^2 \equiv \omega^2/\Delta^2$, $b^2 \equiv 1/a^2$)

$$G_{M_1}^{(1)}(\vec{k}, \omega) = \frac{\omega + \epsilon_{\vec{k}} + i/2\tau_1}{(\omega^2 - \Delta^2 - \epsilon_{\vec{k}}^2)} \left(1 - \frac{-\frac{1}{4\tau_1^2} + \frac{i\omega}{\tau_1} + \frac{b^2}{4\tau_2^2} - \frac{i|b|\Delta}{\tau_2}}{(\omega^2 - \Delta^2 - \epsilon_{\vec{k}}^2)}\right); \ \omega^2 > \Delta^2 \tag{9}$$

$$G_{M_2}^{(1)}(\vec{k},\omega) = \frac{\omega + \epsilon_{\vec{k}} + |a|/2\tau_1}{(\omega^2 - \Delta^2 - \epsilon_{\vec{k}}^2)} \left(1 - \frac{\frac{\omega|a|}{\tau_1} + \frac{a^2}{4\tau_1^2} - \frac{\Delta}{\tau_2} - \frac{1}{4\tau_2^2}}{(\omega^2 - \Delta^2 - \epsilon_{\vec{k}}^2)}\right); \Delta^2 > \omega^2 \tag{10}$$

After using the numerical calculations, $G_{M_{1,2}}(\vec{r})$ can be obtained.

Finite Debye frequency case:

When the Debye frequency is finite, the comparison of the values of $\omega, \Delta$, and $\omega_D$ can produce different $\eta_\omega$, thereby, we can find different Green's functions. By considering the Green's functions in the presence of the non-magnetic impurity, up to the first-order approximation with respect to the inverse collision time, the perturbed Green's function at momentum-frequency coordinate, under assumptions $\omega^2 > \Delta^2$ and $\omega_D < \sqrt{\omega^2 - \Delta^2}$, is obtained as follows

$$G_{NM_2}^{(1)}(\vec{k},\omega) = \frac{\omega}{\tau}\left(\frac{\omega_D(\omega^2 - \Delta^2 + \epsilon_{\vec{k}}^2)}{\pi(\omega^2 - \Delta^2)(\omega^2 - \Delta^2 - \epsilon_{\vec{k}}^2)^2}\right) \tag{11}$$

By doing Fourier transform between momentum and space, which is obtained the perturbed Green's function at space coordinate, we find that the perturbed Green's function under the above conditions does not affect on the concurrence.

Furthermore, under assumptions $\omega^2 > \Delta^2$ and $\omega_D \gg \sqrt{\omega^2 - \Delta^2}$, we obtain

$$G \tag{12}$$

Then, the perturbed Green's function at space coordinate is obtained and again we find that it is not influenced on the concurrence.

Under assumptions $\omega^2 < \Delta^2$ and $\omega_D \gg \sqrt{\Delta^2 - \omega^2}$, we obtain

$$G_{NM_4}^{(1)}(\vec{k},\omega) = -\frac{1}{\tau}\left(\frac{\omega(-\pi\omega_D + 2\sqrt{-\omega^2 + \Delta^2})(\omega^2 - \Delta^2 + \epsilon_{\vec{k}}^2)}{2\pi\omega_D\sqrt{-\omega^2 + \Delta^2}(\omega^2 - \Delta^2 - \epsilon_{\vec{k}}^2)^2}\right) \tag{13}$$

with no effect on the concurrence.

Other assumptions, for example, $\omega^2 < \Delta^2$ accompanied with $\omega_D < \sqrt{\Delta^2 - \omega^2}$, does not show the physical result, because it is in contrast with the statement that expresses always order parameter is less than the Debye frequency.

Now we deal with the rare-earth metal impurity case, and for this case, the perturbed Green's function under assumptions $\omega^2 > \Delta^2$ and $\omega_D < \sqrt{\omega^2 - \Delta^2}$, is obtained as follows

$$G_{M_3}^{(1)}(\vec{k},\omega) = \frac{\omega + \epsilon_{\vec{k}} - \omega_D/\tau_1\pi\omega}{(\omega^2 - \Delta^2 - \epsilon_{\vec{k}}^2)}\left(1 - \frac{\left(\frac{\omega_D}{\tau_1\pi\omega}\right)^2 - \frac{2\omega_D}{\tau_1\pi} + \frac{2\omega_D b^2}{\tau_2\pi} - \left(\frac{\omega_D|b|}{\tau_2\pi\omega}\right)^2}{\omega^2 - \Delta^2 - \epsilon_{\vec{k}}^2}\right) \quad (14)$$

The perturbed Green's function with assumptions $\omega^2 > \Delta^2$ and $\omega_D \gg \sqrt{\omega^2 - \Delta^2}$ is obtained as

$$G_{M_4}^{(1)}(\vec{k},\omega) \approx \left(\frac{1}{\omega^2 - \Delta^2 - \epsilon_{\vec{k}}^2}\right)\left((\omega + \epsilon_{\vec{k}})\left(1 - \frac{\frac{1}{\tau_1}\left(i\omega - \frac{2\omega^2}{\pi\omega_D}\right) + \frac{1}{\tau_2}\left(\frac{2\Delta^2}{\pi\omega_D} - i|b|\Delta\right)}{(\omega^2 - \Delta^2 - \epsilon_{\vec{k}}^2)}\right) + -\frac{\omega}{\pi\tau_1\omega_D} + \frac{i}{2\tau_1}\right) \quad (15)$$

The perturbed Green's function with assumptions $\Delta^2 > \omega^2$ and $\omega_D \gg \sqrt{\Delta^2 - \omega^2}$ is

$$G_{M_5}^{(1)}(\vec{k},\omega) \approx \frac{1}{(\omega^2 - \Delta^2 - \epsilon_{\vec{k}}^2)}\left((\omega + \epsilon_{\vec{k}})\left(1 - \frac{\frac{1}{\tau_1}\left(|a|\omega - \frac{2\omega^2}{\pi\omega_D}\right) + \frac{1}{\tau_2}\left(\frac{2\Delta^2}{\pi\omega_D} - \Delta\right)}{(\omega^2 - \Delta^2 - \epsilon_{\vec{k}}^2)}\right) - \frac{\omega}{\pi\tau_1\omega_D} + \frac{|a|}{2\tau_1}\right) \quad (16)$$

**Acknowledgements**

The authors are grateful for the helpful discussion with Dr. habil. Andreas Osterloh

**Author Contributions**

All authors equally performed the calculations, discussed the results, and participated in writing and reviewing the manuscript. R. A. gives the idea of this research.